\documentclass[epj]{svjour}
\usepackage{graphics}
\newcommand{\beq}{\begin{equation}}
\newcommand{\be}{\begin{eqnarray}}
\newcommand{\eeq}{\end{equation}}
\newcommand{\ee}{\end{eqnarray}}
\newcommand{\ba}{\begin{array}{1}}
\newcommand{\ea}{\end{array}}
\newcommand{\bb}{\begin{thebibliography}}
\newcommand{\eb}{\end{thebibliography}}
\newcommand{\ci}[1]{\cite{#1}}
\newcommand{\bi}[1]{\bibitem{#1}}

\begin{document}

\title{Hadron Multiplicity in Semi-Inclusive
Lepton-Nucleon and Lepton-Nucleus Scattering}

\author{O. Benhar\inst{1} \and S. Fantoni \inst{2} \and G.I. Lykasov \inst{3} 
\and U. Sukhatme \inst{4} \and V.V. Uzhinsky \inst{3} }

\institute{ INFN and Physics Department, 
Universit\`a ``La Sapienza". I-00185 Roma, Italy \and 
International School for Advanced Studies (SISSA). I-34014 Trieste, Italy \and
Joint Institute for Nuclear Research. Dubna, Moscow Region, 141980 Russia \and 
State University of New York at Buffalo. Buffalo, NY 14260-4600, USA }

\date{Received: date / Revised version: date}

\abstract{
We discuss multi-hadron production in both inelastic neutrino-nucleon 
interactions in the current fragmentation region and neutrino-nucleus 
collisions in the target fragmentation region. Our analysis, carried out 
within the framework of the quark-gluon string model, is mainly
focused on the difference between these two processes. We show that the $Q^2$
dependence of hadron multiplicity in the current and target fragmentation 
regions is indeed completely different. The study of inelastic 
$\nu-A$ scattering in the target fragmentation region also provides new 
information on nuclear structure at small $N-N$ distances. The results of
the proposed approach are in satisfactory agreement with the data 
recently obtained at CERN by the NOMAD Collaboration.
}

\maketitle
\section{Introduction}

Deep-inelastic particle interactions play a decisive role in the 
development of the modern theory of fundamental interactions. 
For instance, the investigation of
multi-hadron production in inelastic scattering of leptons off nucleons
and nuclei is a tool to study both the dynamics of such processes and the
quark structure of the target. Of particular interest, in this context, is 
the analysis of the $Q^2$ dependence over a wide range of values. The 
difficulties associated with the increase of
the QCD coupling constant $\alpha_s(Q^2)$ at moderate and low
$Q^2$ have led to the development of alternative approaches to analyze
soft hadron interactions. A well known example is the 
the $1/N$ expansion of the scattering amplitude, $N$ being the number 
of colors or quark flavors, suggested by t'Hooft \ci{thoft} and Veneziano \ci{venez}. 
Based on this approach, the Quark-Gluon String Model
(QGSM) and the Dual Parton Model (DPM) have been developed in
Refs.\ci{qgsm2} and \ci{dpm1,dpm2}, respectively, to analyze hadronic reactions. 
The first application of the QGSM to inelastic lepton-nucleon
processes at moderate and low $Q^2$ has been carried out in Ref.\ci{qgsm1}.

In this paper we investigate inelastic neutrino scattering from both protons and 
nuclei within the QGSM, focusing mainly on the differences between these reactions.
In Section 2 we outline the main features of the theoretical approach, 
whereas Section 3 is devoted to a comparison between our results and the 
data recently obtained at CERN by the NOMAD collaboration. Finally, in 
Section 4 we summarize our findings and state the conclusions.

\section{Theoretical framework}

\begin{figure*}
\resizebox{0.95\textwidth}{!}{\includegraphics{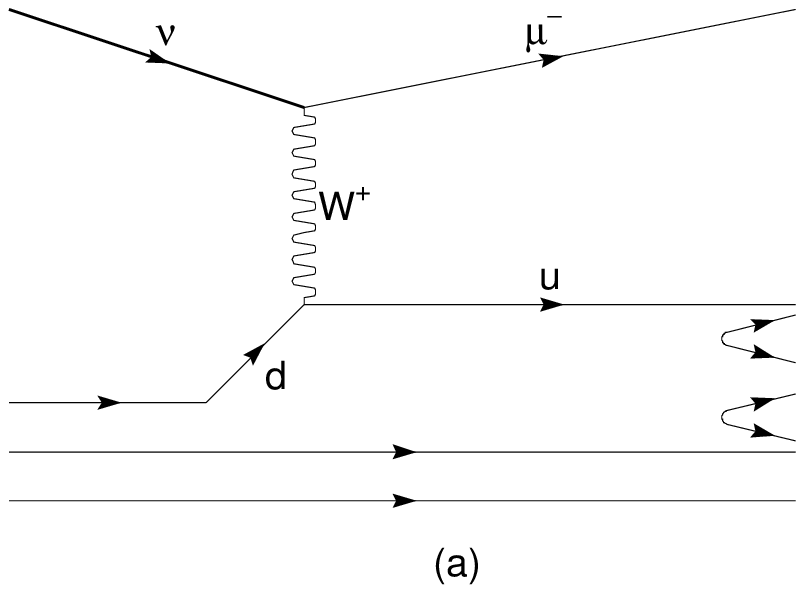}\includegraphics{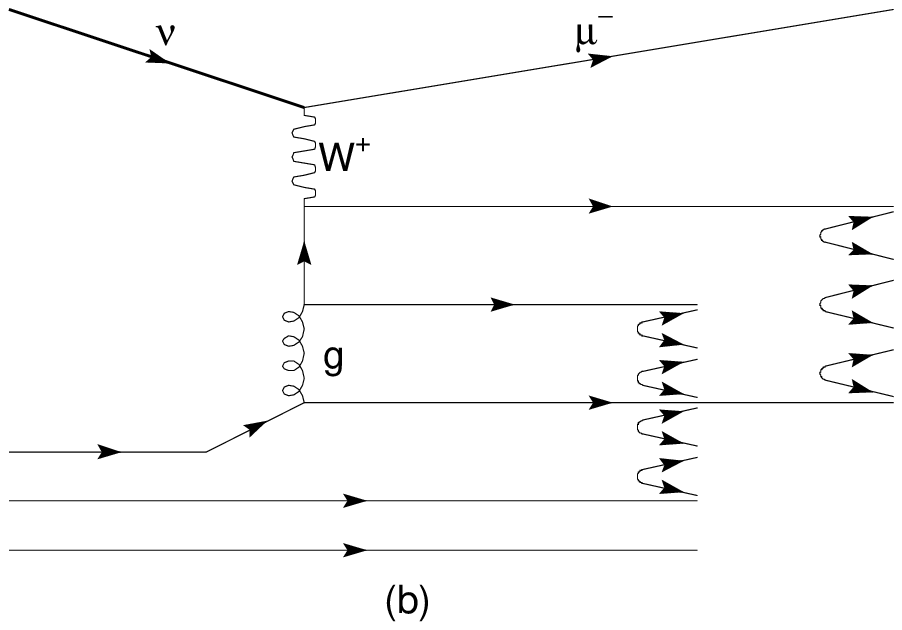}}
\vspace*{-0.85in}
\caption{Planar, one-Reggeon exchange (panel (a)), and 
cylindrical, one-Pomeron exchange (panel (b)), graphs (see text).}
\label{fig1}
\end{figure*}

In the QGSM \ci{qgsm2,qgsm1}, hadron production in the reactions
$\nu(\bar\nu)+p\rightarrow\mu^-(\mu^+)+h+X$ is
described in terms of planar and cylindrical graphs, as shown in
Fig.\ref{fig1}. The planar graph of panel (a) describes neutrino scattering off a
valence quark, corresponding to one Reggeon exchange in the
$t$-channel \ci{qgsm1}, whereas the cylindrical graph of panel (b)
describes neutrino scattering off sea quarks, 
corresponding to one-Pomeron exchange in the $t$-channel
\ci{qgsm1}. The figure also shows the occurrence of hadronization in the
colorless quark-antiquark and quark-diquark strings.

The relativistic invariant distribution of hadrons produced in
the process $\nu(\bar\nu)+p\rightarrow\mu^-(\mu^+)+h+X$
is defined as
\be
\rho_{\nu(\bar\nu)+p\rightarrow\mu^-(\mu^+)+h+X}~=~
E_h\frac{dN}{d^3p_hd\Omega dE^\prime}\ ,
\label{def:dN}
\ee
where $E_h$ and ${\bf p}_h$ are the total energy and three
momentum of the produced hadron, respectively, whereas $E^\prime$ and $\Omega$
are the energy and the solid angle of the final state muon.
The right hand side of Eq.(\ref{def:dN}) can be written in the following general 
form \ci{bfls,lsu}:
\be
\nonumber
\rho_{\nu(\bar\nu)+p\rightarrow\mu^-(\mu^+)+h+X} & = & 
\Phi(Q^2)\left\{F_P(x,Q^2;z,p_{ht}) \right.  \ \ \ \ \ \ \ \ \ \  \\
&  & \ \ \ \ \ \ \ \ \  \left. + \ F_C(x,Q^2;z,p_{ht})
\right\} \ ,
\label{def:rho}
\ee
with
\beq
\Phi(Q^2)~=~mE\frac{G^2}{\pi}\frac{m_W^2}{Q^2+m_W^2}\ ,
\eeq
where $G$ is the Fermi weak coupling constant, $E$ is the energy
of the incoming neutrino, $m$ and $m_W$ are the nucleon and 
$W$-boson masses, respectively, $x=Q^2/2(p_\nu\cdot k)$ is
the Bjorken variable and $p_\nu$ and $k$ are the four momenta of
the initial neutrino and nucleon. 
In general, $z$ is the light cone variable defined as
$z=(E_h+p_{hz})/(E+p_z)$, where $p_{hz}$ is the component of the 
hadron momentum
parallel to the momentum of the incoming neutrino, $p_z$, (see, e.g.,
Refs.\ci{qgsm2,qgsm1} and \ci{dpm1,dpm2}).
At large energies of the final hadron it                                       
reduces to the longitudinal momentum fraction of the hadron, with 
respect to the neutrino,
in the rest frame of the target proton. The variable $z$ can also be 
treated as the Feynman variable $x_F=2p_L^*/W_X$, defined as the 
longitudinal momentum fraction in the hadronic center of mass system (HCMS).
Here $p_L^*$ denotes the longitudinal
hadron  momentum in the HCMS and $W_X$ is the mass of the hadrons
produced in the reaction. Finally, $p_{ht}$ is the transverse momentum
of the produced hadron with respect to the current (hadronic jet)
direction, (see, e.g. Ref.\ci{nomad2}).

The probabilty distributions of hadron production associated with
the planar and cylindrical graphs of Fig.\ref{fig1} are 
given by $F_P(x,Q^2;z,p_{ht})$ and $F_C(x,Q^2;z,p_{ht})$, respectively.
They can be computed
either analytically, as in Ref.\ci{bfls}, or using the Monte Carlo (MC) 
approach to generate all quark-antiquark and quark-diquark strings shown
in Fig.\ref{fig1}, as done in Ref.\ci{lsu}.

The main ingredients for the calculations of observables in the
reaction under discussion are the quark distributions in a nucleon and
their fragmentation functions to hadrons. In addition to the dependence 
upon $x$, quark
distributions also depend on $Q^2$ and the transverse momentum
$k_t$. Following Ref. \ci{bfls}, we use a
factorized form for these distributions:
\be
q_f(x,Q^2;k_t)~=~ q_f(x,Q^2)g_q(k_t)\ ,
\label{def:xqkt}
\ee
with the function $g_q$ chosen in the form
\be
g_q(k_t)~=~\frac{B^2}{2\pi}e^{-Bk_t}
\label{def:gkt}\ ,
\ee
where $B=1/\langle k_t\rangle \simeq 4 (GeV/c)^{-1}$, while
$\langle k_t\rangle \simeq 0.25$ GeV/c is the average transverse
momentum of a quark in a nucleon.
As for the function $q_f(x,Q^2)$, we use
the fit suggested in Ref.\ci{ckmt}, including true
Regge $x$-asymptotic at $x\rightarrow 0,
x\rightarrow 1$ and small $Q^2$, and its QCD prediction
at large $Q^2$.

In general, the fragmentation functions (FF) of quarks
(diquarks) into hadrons, $D_{q(qq)}^h$,  depend on the hadron
momentum fraction $z_1$ and the hadron transverse momentum
with respect to a quark (diquark) momentum
direction, ${\widetilde p}_{ht}$. Here we choose again the factorized form
\be
D_{q(qq)}^h(z_1,{\widetilde p}_{ht})~=~ D_{q(qq)}^h(z_1)g_q({\widetilde p}_{ht})\ ,
\label{def:Dq}
\ee
with the function $g_q$ defined as in Eq.(\ref{def:gkt}).

The functions $D_{q(qq)}^h(z_1)$ are obtained, according to
the recursive cascade model procedure suggested in \ci{ff1}, from the 
integral equation 
\be
D_{q(qq)}^h(z_1)~=~
f(z_1)~+~\int_{z_1}^1 \frac{dx}{x} f(x) D_{q(qq)}^h\left(\frac{z_1}{x} \right) \ ,
\label{def;frf}
\ee
the function $f(x)$ being chosen in the form
\be
f(x)~=~x^\beta(1-x)^\gamma\ .
\label{def:fx}
\ee
According to the main assumptions underlying the QGSM, the 
fragmentation functions
$D_{q(qq)}^h(z_1)$ should satisfy true Regge asymptotic
at $z_1\rightarrow 1$ and $z_1\rightarrow 0$ \ci{qgsm2}.
These constraints determine the values of the parameters.
The detailed procedure is presented in Ref.\ci{shmuzh}.
In general the FF depend not only on $z_1$ and $p_{ht}$,
but also on $Q^2$. At low $Q^2$ they have to reproduce the true
Regge asymptotic \ci{qgsm2}, while  at large $Q^2$ they have to describe
$e^+e^-$ annihilation data. As we mainly analyze inelastic $\nu-N$ interactions 
at moderate $Q^2$, one can assume that
the $Q^2$-dependence of the FF be negligibly weak.

Let us now consider pion production in the target
fragmentation region in $\nu-A$ interactions.
Our study of these processes focuses on the kinematical region of
large pion momentum ($p_\pi>0.3$ GeV/c), where the effects of final state
interactions (FSI), leading to pion absorption associated with production
of baryon resonances, are expected to be small \ci{frst,al} and the impulse
approximation (IA) cane be safely used.

Within the framework of the IA
the relativistic invariant semi-inclusive spectrum of hadrons, in
particular pions, produced by the semi-inclusive inelastic process
$\ell +  A \rightarrow  \ell^\prime + \pi + X$,
\beq
\rho_{\ell +  A \rightarrow  \ell^\prime + \pi + X} \equiv 
E_hd\frac{\sigma}{d^3p_h d\Omega dE^\prime} \ ,
\eeq
can be written in the convolution form \ci{bfls}
\be
\nonumber
& & \rho_{\ell + A \rightarrow \ell^\prime + \pi + X}(x,Q^2;z,p_t) = 
\int_{z\leq y} dyd^2k_t
f_A(y,Q^2,k_t)  \ \  \\  
& & \ \ \ \ \times 
\left[ \frac{Z}{A}\rho_{\ell + p \rightarrow \ell^\prime + \pi + X}(x/y,Q^2;z/y,p_t-k_t) 
\right. \\  \nonumber
& & \ \ \ \ \ \ \ \ \ \ \ \ \ \ + \left. 
\frac{N}{A}\rho_{\ell + n \rightarrow \ell^\prime + \pi + X}(x/y,Q^2;z/y,p_t-k_t) \right] \ ,
\label{def:lasp}
\ee
where $Z$, $N$ and $A$ are the numbers of nucleons, protons and neutrons
in the nucleus, $z=(p p_\nu)/(P_A p_\nu)M_A/m$,
$p$ is the four-momentum of the produced
pion and $M_A$ and $m$ are the nucleus and nucleon masses, respectively.
$\rho_{\ell + p(n) \rightarrow \ell^\prime + \pi + X}$ denotes the semi-inclusive 
spectra of pions produced by interactions of the lepton $\ell$ with a quasi-free
proton (neutron). The nucleon distribution function
$f_A(y,k_t)$ is defined as \ci{bfls1}
\be
f_A(y,k_t) = \int dk_0dk_z S(k)y\delta\left(y-\frac{M_A}{m}
\frac{(k q)}{(P_A q)}\right)\ ,
\label{def:fa}
\ee
where $q$ is the four-momentum transferred by the lepton, $q^2=-Q^2$,
$S(k)$ is the relativistic invariant function describing the
nuclear vertex with an outgoing virtual nucleon,
$y=(M_A/m)(kp_\nu)/(P_Ap_\nu)$ and
$P_A$ ,$k$ and $p_\nu$ are the four-momenta of the nucleus, nucleon and initial
neutrino, respectively.

The distribution function $f_A(y)$, defined as the integral
of $f_A(y,k_t)$ over $d^2k_t$, can be calculated within nuclear 
many-body theory approximating $S(k)$ with the nonrelativistic
spectral function $P(k,E)$, yielding the probability of finding
a nucleon with momentum $\bf k$ and removal energy $E=m-k_0$
\ci{bff}. However, due to the limited range of momentum and
removal energy covered by nonrelativistic calculations of
$P(k,E)$ (typically $|{\bf k}|<k_{min}\sim 0.7-0.8$ GeV/c
and $(m-k_0) < 0.6$ GeV, see e.g. \ci{bff}), this procedure
can only be used in the region $y < y_0 \sim 1.7-1.85$. An
alternative approach to obtain $f_A(y)$ at larger $y$, based
on the calculation of the overlap of the relativistic invariant
phase-space available to quarks belonging to strongly correlated
nucleons, has been proposed in Refs.\ci{bfls1,bfl3}.

The convolution form of Eq.(\ref{def:lasp}) implies that
to analyze the production of pions emitted in the whole backward
hemisphere by $\nu-A$ interaction one needs
to know the dependence of the nucleon distribution
upon $y$ and $k_t$. We assume the following factorized form:
\be
f_A(y,k_t)=f_A(y)g_A(k_t)\ ,
\label{def:faykt}
\ee
where the function $g_A$ is chosen to be a Gaussion 
\be
g_A(k_t)=\frac{1}{\pi \langle k_t^2 \rangle}\exp(-k_t^2/\langle k_t^2 \rangle)\ ,
\label{def:ga}
\ee
$\langle k_t^2 \rangle$ being the average value of the squared transverse
momentum of a nucleon in  the target nucleus.


\section{Comparison to data}

\begin{figure}
\resizebox{0.45\textwidth}{!}{\includegraphics{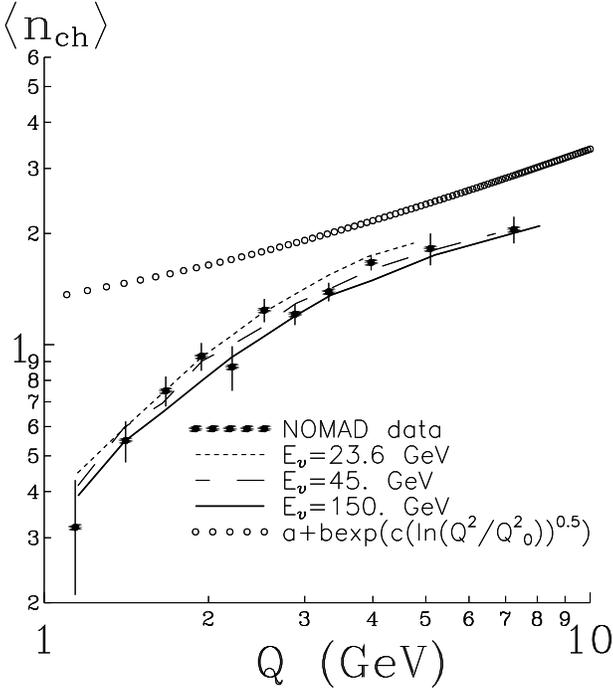}}
\caption{
Mean multiplicity of charged hadrons in the current
fragmentation region for the process $\nu+p\rightarrow\mu^-+h+X$,
as a function of the momentum transfer $Q$.
The open circles correspond to the QCD fit given by eq.(\protect\ref{def:nchqcd}),
while the solid, long-dash and short-dash lines correspond to our calculations at
$E_\nu=$ 150, 45 and 23.6 GeV, respectively.
The experimental data are taken from Ref.\ci{nomad1}
}
\label{fig2}
\end{figure}

\begin{figure}
\resizebox{0.45\textwidth}{!}{\includegraphics{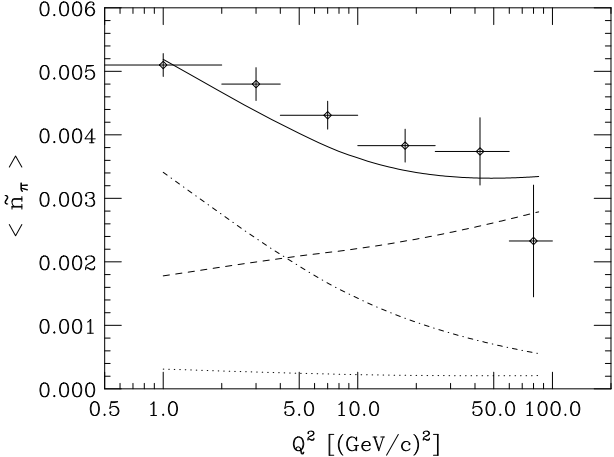}}
\caption{Mean multiplicity of charged pions produced in the
backward hemisphere in $\nu-C$ interaction as a function of $Q^2$.
The solid line shows the results of the full calculation, including
both graphs of Fig.\ref{fig1}. The dashed and dash-dot lines correspond to the
separated contributions of the cylindrical graph of Fig.\ref{fig1} (b) and the
planar graph of Fig.\ref{fig1} (a), respectively. The dots show the results
obtained setting $P({\bf k},E)\equiv 0$ in the high energy-momentum domain, not 
covered by the nonrelativistic calculation of Ref.\ci{bff}. The experimental data are taken
from Ref.\protect\ci{nomad2}.}
\label{fig3}
\end{figure}

\begin{figure}
\resizebox{0.45\textwidth}{!}{\includegraphics{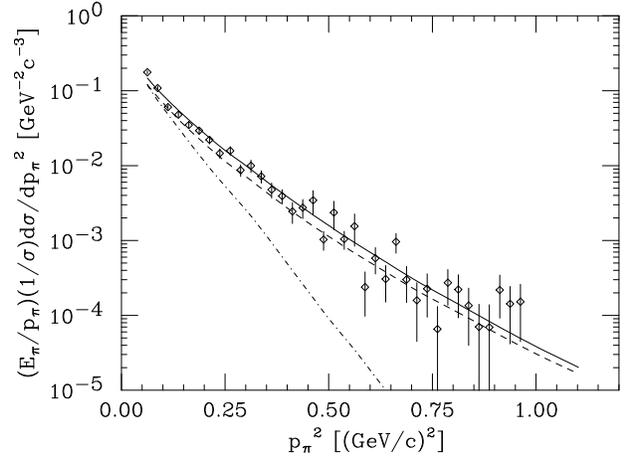}}
\caption{
$p_\pi^2$ -dependence of the spectrum of backward pions produced in inelastic
$\nu-C$ interaction. The solid curve corresponds to the full calculation whereas 
the dashed line has been obtained including only the cylindrical graph of 
Fig.\ref{fig1} (b). The dash-dot line shows the results obtained setting 
$P({\bf k},E)\equiv 0$ in the high 
energy-momentum domain, not covered by the nonrelativistic calculation of 
Ref.\ci{bff}. The experimental data are taken from
Ref.\protect\ci{nomad2}.}
\label{fig4}
\end{figure}

The results of our calculation of different observables of the 
reactions $\nu+N\rightarrow\mu^-+h+X$ and
$\nu+C^{12}\rightarrow\mu^-+\pi^-+X$ are presented in Figs.\ref{fig2}$-$\ref{fig4}.

The mean charged multiplicity in the current region
is shown in Fig.\ref{fig2}. The multiplicity measured by the NOMAD 
collaboration \ci{nomad1}, $\langle n_{ch} \rangle$, turns out to be close to 
$\langle n_{ch} \rangle/2$
measured in $e^+e^-$ experiment at energy $E=\sqrt{s}$
and to $\langle n_{ch} \rangle$ obtained from $ep$ and ${\bar\nu}_\mu p$ at $E=Q$.
In Fig.\ref{fig2} we compare our results to $\langle n_{ch}^{QCD} \rangle/2$ 
obtained from 
a QCD calculation of the charged multiplicity in $e^+e^-$. The QCD results,  
represented by open circles, correspond to 
evolution of parton distributions in the leading log approximation, 
yielding the following fit for 
$\langle n_{ch}^{QCD} \rangle$ (see Ref.\ci{nomad1}):
\be
n_{ch}^{QCD}~=~a+b\,\exp \left( c\sqrt{\log(Q^2/Q_0^2)}\right)\ ,
\label{def:nchqcd}
\ee
where
$a=2.257$, $b=0.094$, $c=1.775$ and $Q_0=1$ GeV/c. The solid, long-dash 
and short-dash lines in Fig.\ref{fig2} correspond to our Monte Carlo
calculations at initial neutrino energies $E_\nu=150$  GeV,
$E_\nu=45$ GeV, $E_\nu=23.6$ GeV, respectively. The NOMAD
experimental data are averaged over the initial energy
\ci{nomad1}. 

Fig.\ref{fig2} shows that the QCD fit, while failing to 
reproduce the NOMAD data on the charged multiplicity
at $Q < 5$ GeV/c, provides a reasonable description of the behavior at 
large $Q$. As discussed in \ci{nomad1} the same fit also satisfactorily describes
$\langle n_{ch} \rangle$ obtained from $e^+e^-$, $ep$ and ${\bar\nu}_\mu p$
experiments. On the other hand, the approach proposed in Ref.\ci{lsu} 
describes the NOMAD data at $Q~<~5$ GeV/c rather well.

The multiplicity of pions, normalized to the cross section of the
process $\nu+A\rightarrow\mu+X$, $\sigma$, is defined as
\be
\nonumber
\langle {\widetilde n}_\pi \rangle & \equiv & 
\frac{\langle n_\pi \rangle}{\sigma} =
\frac{1}{\sigma}\int_{x_{min}}^{x_{max}}dx \\ 
  & \times & \int_{z_{min}}^{z_{max}}
\frac{dz}{z}\int_0^{p_{max}}d^2p\ \rho_{\nu+A\rightarrow\mu+\pi+X}\ .
\label{def:multpi}
\ee
It can be split in two parts, corresponding to the planar,or
one-Reggeon exchange, diagram (Fig.\ref{fig1} (a)) and cylindrical, or
one-Pomeron exchange graph (Fig.\ref{fig1} (b)).

The NOMAD collaboration carried out a study of inelastic
$\nu-C$ interactions in which the negative pions emitted in
the whole backward hemisphere, with respect to the incoming
neutrino beam, were detected \cite{nomad2}. The data of Fig.\ref{fig3}
show the nultiplicity of negative pions carrying momenta
$0.35<p_\pi<0.8$ GeV/c, measured in a kinematical setup in
which $W_X$ increases as $Q^2$ increases. Theoretical calculations
have been performed applying the same kinematical conditions.

As shown in Fig.\ref{fig3}, the planar graph provides the main
contribution at small $Q^2$, while the cylindrical one dominates
at $Q^2>10$ (GeV/c)$^2$. It clearly appears that both diagrams have
to be included to explain the observed $Q^2$ dependence.
The dotted line in Fig.\ref{fig3} has been obtained setting $P({\bf k},E)=0$
in the domain of large energy and large momentum, not covered by
the nonrelativistic calculations of ref.\ci{bff}. Comparison between
the solid and dotted line shows that the dominant contribution to
$<{\widetilde n}_\pi>$ does indeed come from the high momentum tail of the nucleon
distribution.

In Fig.\ref{fig4} we compare the $p_\pi^2$ dependence of the experimental
spectrum to the one resulting from our approach. 
The contribution of the cylindrical graph dominates the spectrum
and the inclusion of the high momentum tail of the nucleon
distribution, corresponding to $p>0.4$ GeV/c, is clearly needed to describe
the data.

\section{Summary and conclusions}

In this paper we have analyzed inelastic $\nu-N$ and
$\nu-A$ processes at low and moderate $Q^2$ within the QGSM.
The main conclusions can be summarized as follows. The dynamics of hadron
production in the current fragmentation and target fragmentation regions
is different. At low $Q^2$ the contribution of the cylindrical graph
to the hadron multiplicity is much more prominent in the first kinematical region
than it is in the second one. 
There is some analogy between pion production
in the backward hemisphere from target fragmentation and soft
$h-N$ processes at $x_F\rightarrow -1$, where the one-Pomeron exchange
graph dominates at large initial energy (see, e.g., Ref.\ci{qgsm1}).
In conclusion, the contributions of the graphs of Figs.\ref{fig1} (a) and (b)
have different $Q^2$ dependences in the different kinematical
regions corresponding to current and target fragmentation.

The conventional perturbative QCD calculations of ha\-dron
multiplicities, both in $\nu-N$ and $\nu-A$ inelastic interactions, 
do not reproduce the NOMAD data, whereas the application of the QGSM
allows for a rather satisfactory description.
As for the role of nuclear structure, the dominant contribution
to the spectrum of pions emitted in the backward hemisphere in $\nu-A$
inelastic scattering is clearly coming from the high momentum tail of the
nucleon distribution, which can be described in terms of overlaps of
distributions of three-quark colorless objects \ci{bfls2}.

\end{document}